\documentclass[%
 reprint,
superscriptaddress,
 amsmath,amssymb,
 aps,
 prl,
]{revtex4-2}

\usepackage[normalem]{ulem}
\usepackage{graphicx}
\usepackage{dcolumn}
\usepackage{bm}
\usepackage{hyperref}

\usepackage[dvipsnames]{xcolor}

\newcommand{\sect}[1]{\textit{#1}---}

\begin{document}

\preprint{APS/123-QED}

\title{
        Steady reservoir-mediated
        self-trapping of 
        polariton condensates}

\author{Pavel N. Kozhevin}
\affiliation{Abrikosov Center for Theoretical Physics, MIPT, Dolgoprudnyi 141701, Russia}
\affiliation{Spin Optics Laboratory, Saint-Petersburg State University, St. Petersburg, 198504, Russia}

\author{Oleg I. Utesov}
\affiliation{Department of Physics, Korea Advanced Institute of Science and Technology (KAIST), Daejeon 34141, Republic of Korea}
\affiliation{Center for Theoretical Physics of Complex Systems, Institute for Basic Science (IBS), Daejeon 34126, Republic of Korea}

\author{Igor S. Aranson}
\affiliation{Departments of Biomedical Engineering, Chemistry, and Mathematics, The Pennsylvania State University, University Park 16802, USA}

\author{Sergey V. Koniakhin}
\affiliation{Russian Quantum Center, Skolkovo, Moscow, 121205, Russia}

\author{Anton V. Nalitov}
\affiliation{Abrikosov Center for Theoretical Physics, MIPT, Dolgoprudnyi 141701, Russia}
\affiliation{Russian Quantum Center, Skolkovo, Moscow, 121205, Russia}

\date{\today}

\begin{abstract}
    Nonequilibrium bosonic condensates of exciton-polaritons generated by tightly focused incoherent optical beams are typically expected to form a stationary ballistically expanding state with the notable exception of experimentally observed compact self-trapped states [Phys. Rev. Lett. 123, 047401 (2019)].
    We show that the back-action of the condensate on the optically pumped excitonic reservoir via bosonic stimulated scattering can cause local reservoir depletion, producing a steady condensate confined in a self-induced potential trap.
    Notably, this self-trapped state is formed by persistent converging rather than radially expanding polariton currents and exists even below the conventionally defined condensation threshold.
    This state remains steady over nanosecond-scale times and is eventually ejected from the self-induced trap due to a bullet-type instability, enabling spiking neuromorphic dynamics in networks of polariton condensates.

\end{abstract}

\maketitle

Nonequilibrium exciton-polariton condensates present an intrinsically two-component system, where a coherent superfluid fraction coexists with an incoherent reservoir of uncondensed polaritons, sharing a similarity with superfluids near the lambda transition \cite{Kavokin2007,Carusotto2013}.
In this picture, the reservoir simultaneously provides the condensate with gain, compensating losses by stimulated scattering \cite{Haug2014}, and affects its dynamics via repulsive interactions \cite{Wouters2007,Ferrier2011}.
Controlling the spatial profile of the nonresonant pump therefore allows engineering effective non-Hermitian potential landscapes for the condensate~\cite{Schneider2017} and producing steady dissipative soliton states \cite{Ostrovskaya2012,Hu2026}.

One facet of the condensate-reservoir interplay is revealed by ballistically expanding condensates, which harness the potential energy of repulsion from the reservoir induced by a focused pump spot and convert it into kinetic energy of radial expansion beyond the pump spot \cite{Christmann2012}.
This ballistic expansion has been employed for optically controlled dissipative coupling of adjacent condensates \cite{Ohadi2016,Alyatkin2020}, leading to graphs \cite{Berloff2017} and lattices \cite{Alyatkin2021} of mutually coherent condensates.
The latter simulate complex solid-state physics such as quasi-crystals \cite{Alyatkin2025} and geometric frustration \cite{Cookson2021,Alyatkin2024APL}.

In contrast, ring-shaped pumping exploits the same repulsive potential to confine an extremely coherent condensate inside an all-optical trap \cite{Askitopoulos2013,Orfanakis2021}, enabling controllable switching of confined states \cite{Askitopoulos2015,Sun2018}.
Optically trapped nonequilibrium condensates support spin bistability \cite{Ohadi2015,Pickup2018}, while lattices of such traps realize spin ordering \cite{Ohadi2017} and can have nontrivial band topology \cite{Sigurdsson2019,Pickup2020,Pieczarka2021}.
Larger annular traps accommodate persistent vortex currents \cite{Dreismann2014,Liu2015,Alyatkin2024} and have recently been demonstrated as a promising platform for supersolidity \cite{Kozhevin2025}.

Intriguingly, under similar experimental conditions of a single continuous-wave (CW) Gaussian pumping spot, formation of a self-trapped rather than the expected ballistically expanding condensate has been reported \cite{Ballarini2019}.
Observations of such self-trapping under both CW and pulsed pumping \cite{Dominici2015} were attributed to local crystal lattice heating and the formation of a collective polaron, despite the absence of direct experimental evidence of heat release.
At the same time, single-shot experiments with pulsed non-resonant excitation have revealed sporadic self-trapping caused by local depletion of the reservoir by the condensate \cite{Estrecho2018,Bobrovska2018}.
This raises a natural question: can the same mechanism of local reservoir depletion \cite{Wouters2007}, forming a potential trap and thus suppressing ballistic expansion, be responsible for stable condensate self-trapping under CW focused optical pumping and its real-space collapse in the pulsed pumping regime?

\begin{figure}
    \centering
    \includegraphics[width=1.0\linewidth]{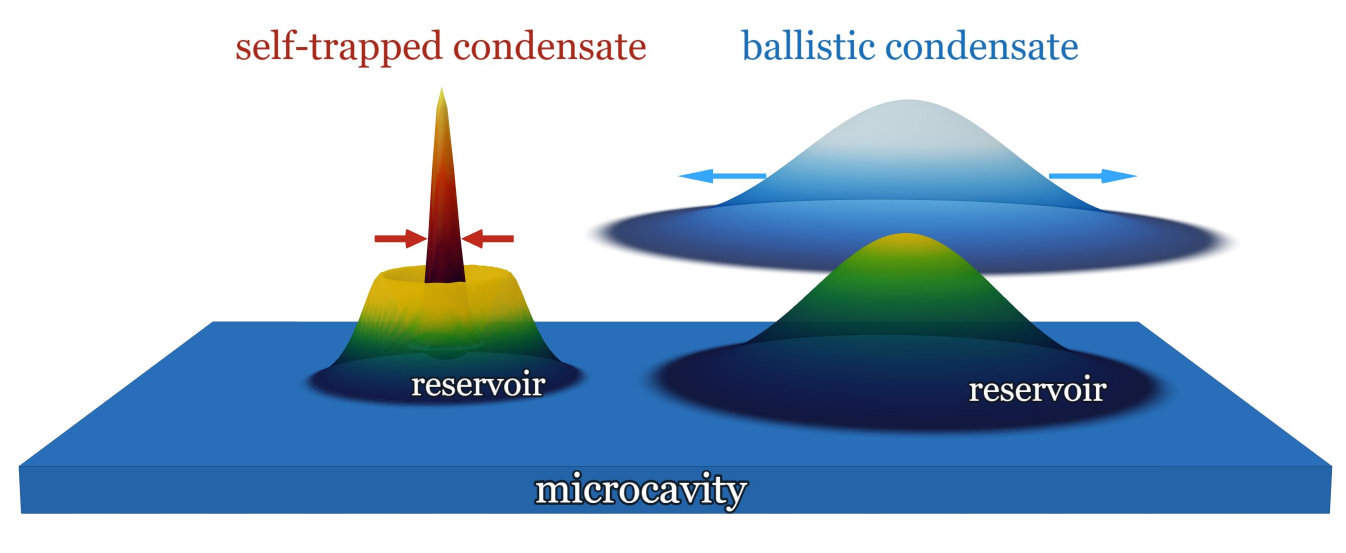}
    \caption{The two types of steady condensate-reservoir states. Left: the self-trapped condensate (red) residing in a self-induced potential trap due to the hole-burning effect. Right: the standard ballistic condensate (blue) expanding due to the reservoir-induced potential hump. Arrows indicate the direction of polariton current density.}
    \label{fig:1}
\end{figure}

In this Letter, we demonstrate that the standard extended Gross-Pitaevskii Equation (eGPE) model accounting for coupled dynamics of the coherent condensate and incoherent reservoir, accommodates both ballistic expansion and self-trapping regimes in a unified picture, schematically illustrated in Fig. \ref{fig:1}.
In stark contrast to the conventionally expected steady ballistically expanding condensate \cite{Ostrovskaya2012}, stationary self-trapped condensates exhibit converging polariton currents, indicated by arrows in Fig.~\ref{fig:1}, flowing inward due to a self-induced potential trap.
This compact self-consistent steady state can also be viewed as resulting from the effective reservoir-mediated condensate attraction, which emerges from the interplay between the local reservoir depletion and the condensate-reservoir repulsion.
Notably, the self-trapped condensate can coexist with the trivial no-condensate state below the condensation threshold predicted by the linear model, giving rise to an on-off bistability regime.
A more detailed numerical analysis reveals weak instability of the self-trapped state and the emergence of a previously overlooked non-adiabatic regime of persistent condensate relaxation oscillations.
Finally, we show that our findings are fully consistent with existing experimental observations of steady compact self-trapped condensates under focused CW pumping  \cite{Ballarini2019} and the real-space condensate collapse under pulsed excitation \cite{Dominici2015}.

We use the eGPE model for the coupled dynamics of the condensate $\Psi$ and the reservoir density $n$:
\begin{subequations}\label{eq:eGPE}
\begin{align}
    i \hbar \partial_t \Psi =& \left[ -{\hbar^2 \over 2m^*} \nabla_r^2 + {\alpha + i \beta \over 2} n + \alpha_1 |\Psi|^2 - i\hbar {\Gamma \over 2}\right]\Psi, \label{eq:eGPEcond} \\ 
    \partial_t n =& P - \left( \gamma + {\beta \over \hbar} |\Psi|^2 \right)n, \label{eq:eGPEres}
\end{align}
\end{subequations}
where $m^*$ is the polariton effective mass, $\alpha>0$ and $\beta>0$ are the Hermitian and anti-Hermitian parts of the condensate-reservoir interaction, $\alpha_1>0$ is the condensate self-interaction strength, $P(\mathbf{r})$ is the nonresonant pump power, $\Gamma$ and $\gamma$ are the condensate and reservoir inverse lifetimes.
Using the characteristic time $\Gamma^{-1}$ and size $L = \sqrt{\hbar/(m^*\Gamma)}$, we switch to dimensionless variables $\tau = t\Gamma/2$, $\boldsymbol{\rho} = \boldsymbol{r} /L$, $\eta = n\beta / (\hbar \Gamma)$, $\psi = \Psi \sqrt{2 \beta / (\hbar \Gamma)}$, and 
dimensionless parameters $\varepsilon = \alpha / \beta > 0$, $\xi = \alpha_1 / \beta >0$, and $a = 2\gamma / \Gamma>0$, $p = 2 P\beta / (\hbar \Gamma^2)$.
While parameters $\varepsilon$ and $\xi$ quantify the degree of Hermiticity of the condensate repulsion from the reservoir and from itself, parameter $a$ governs the intrinsic nonadiabaticity of the two-component system through the ratio of the two decay rates.
We specifically consider the case of Gaussian pumping $p = p_0 \exp(-\rho^2/R^2)$.


Stationary polariton modes below or at the condensation threshold ($|\psi|^2=0$) are described within the \emph{linear approximation}, in which the reservoir density reads $\eta(\rho) = p(\rho)/a $.
Near the center of the pumping spot $\rho = 0$ we use the parabolic approximation for the reservoir density $\eta(\rho) = \eta_0 + \eta_2 \rho^2$ with $\eta_0 = p_0/a$ and $\eta_2 = -p_0/(aR^2)$, which yields a non-Hermitian quantum harmonic oscillator with complex energies $E_{n_x,n_y}$ (see Ref.~\cite{Utesov2025} for details).
The condensation condition for the dominant mode $\mathrm{Im}\lbrace E_{0,0} \rbrace = 0$ yields the threshold pumping power $p_\text{th} = a \left( I + \sqrt{I^2 + 1} \right)^2$ with $I = \mathrm{Im \{ {i} \sqrt{(\varepsilon+i)}\}}/R$.
The corresponding condensate wavefunction reads $\psi(\rho) = \psi_0 \exp [ - s\rho^2/2 ]$ with $s=- i \sqrt{p_\text{th}(\varepsilon+i)/(aR^2)}$.
Its characteristic size reads $R_c(\varepsilon,R) = \mathrm{Re}\left\{s\right\}^{-1/2}<R$, meaning that the condensate is smaller in size than the pump spot at the initial stage of formation from a fluctuation seed, which is governed by the linear approximation (see End Matter).

This effect is further amplified by the effective reservoir-mediated attraction in the \emph{weakly nonlinear regime} ($|\psi|^2\ll a$).
Indeed, expanding the stationary reservoir density under the adiabatic approximation to the first order in low condensate density as $\eta \approx p/a - p|\psi|^2/a^2$, one obtains the attractive correction $\xi^\prime = -\varepsilon p /a^2 $ to the condensate repulsive self-interaction $\xi$.
The net effective interaction is pumping-dependent and can be attractive above the linear condensation threshold provided $\varepsilon p_\text{th}/a^2 > \xi$.
The asymptotic condensation threshold divergence for small pump spot sizes $p_\text{th}\sim 1/R^2$ indicates that the attraction mechanism is dominant over size-independent repulsion $\xi$ for tightly focused pump spots (see End Matter).


Attractive cubic interaction nonlinearity is known to lead to condensate instability resulting in its real-space collapse \cite{Chiao1964}.
Here, such a collapse is avoided due to local depletion of the reservoir density by the condensate, which saturates the effective attraction strength.
This effect is captured within the nonadiabatic two-component model beyond the weakly nonlinear approximation.
Keeping the ansatz $\psi(\rho,\tau) = \sqrt{n_0(\tau)} \exp [ - (w(\tau)-ib(\tau))\rho^2/2 ]$ for the condensate and $\eta(\rho,\tau) = \eta_0(\tau) + \eta_2(\tau)\rho^2$ for the reservoir density, we obtain the dynamical system
\begin{align}
    \dot{n_0} =& 2n_0(\eta_0 - 1 - 2b), \quad \dot{w} = - 4wb - 2\eta_2,  \nonumber \\
    \dot{b} =& 2(w^2 - b^2 - \varepsilon \eta_2 + \xi n_0 w). \label{eq:dyn_cond}
\end{align}
for the condensate variables, supplemented with dynamical equations for those of the reservoir:
\begin{equation}
    \dot{\eta_0} = p_0 - (a+n_0)\eta_0, \,
    \dot{\eta_2} = -{p_0 \over R^2} - (a+n_0) \eta_2 + n_0 w \eta_0. \label{eq:dyn_res}
\end{equation}


\begin{figure}
    \centering
    \includegraphics[width=1.0\linewidth]{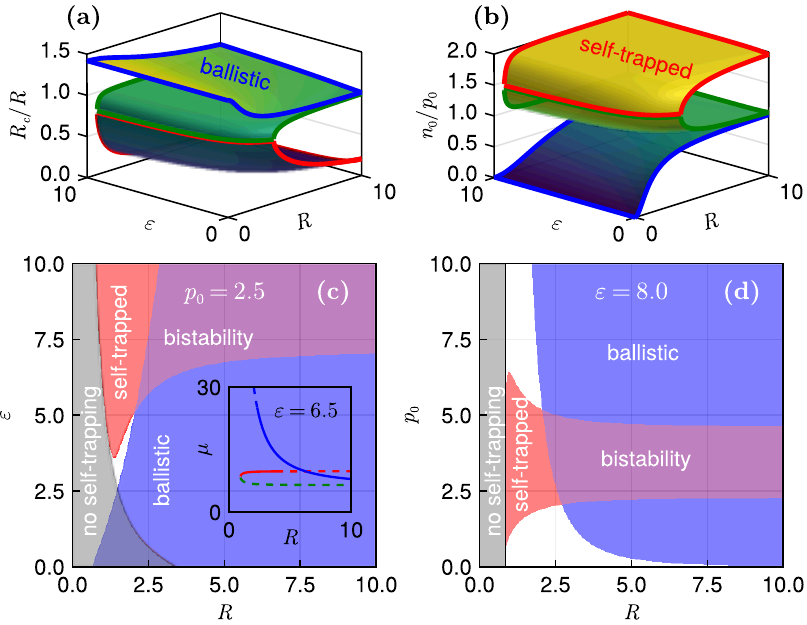}
    \caption{Analytic results in the deeply nonlinear regime. a) The ratio of the condensate and the pump spot radii and b) the maximal condensate density, normalized by the pumping power for the self-trapped ($R_c<R$, red), the intermediate (green), and the ballistic ($R_c>R$, blue) solutions of Eqs. (\ref{eq:dyn_cond},\ref{eq:dyn_res}) at $a=\xi=0$. The stability diagrams on the (c) $R-\varepsilon$ ($p_0=2.5$) and the (d) $R-p_0$ ($\varepsilon=8$) planes show bistability (purple), absence of any stable fixed points (white), and the absence of self-trapped states (gray).
    Inset: condensate energy $\mu$ as a function of $R$ for $\varepsilon=6.5$, showing that the self-trapped (red) state is more energetically favorable for small pump spot sizes.}
    \label{fig:2}
\end{figure}

Although in the most general case the system (\ref{eq:dyn_cond},\ref{eq:dyn_res}) is not analytically tractable, useful insights can be drawn from the \emph{deeply nonlinear regime} $|\psi|^2 \gg a$ (see End Matter for details).
In particular, the system has three stationary solutions whose analytically calculated properties are illustrated in Fig. \ref{fig:2}.
One of them corresponds to an extended condensate $R_c>R$, is referred to as ballistic, and is shown in blue.
Although the other two are both characterized by $R_c<R$, the wider, shown in green, is dynamically unstable.
The smallest in size self-trapped state is at the same time characterized by the highest maximal density $n_0$, as shown in Figs. \ref{fig:2}a,b.
In stark contrast to the conventional ballistic state, the self-trapped condensate is characterized by centripetal polariton currents due to a self-induced effective potential trap, similar to those induced by annular pump beams \cite{Nalitov2019}.

While the relative size $R_c/R$ and normalized peak density $n_0/p_0$ only depend on the pump spot size $R$ and interaction parameter $\varepsilon$ within this approximation for both types of states, their stability deduced from Lyapunov exponents of the Jacobi matrix depends on the peak pump power $p_0$ as well.
Phase diagrams in Figs. \ref{fig:2}c,d show that the self-trapped state can be uniquely stable for sufficiently small pump spots close to the limiting size $R_\text{crit}(\varepsilon)$, below which no self-trapped fixed point exists (gray area).
For wider pump spots, self-trapped condensates are either unstable or bistable with the trivial ballistic state, in qualitative agreement with the dominance of the effective reservoir-mediated attraction for small pump spots, following from the weakly nonlinear analysis.
The energy of the condensate described by Eqs. (\ref{eq:dyn_cond},\ref{eq:dyn_res}) is given by $\mu = 2w + \varepsilon \eta_0 + \xi n_0$.
However, if the self-energy $\xi n_0$ is negligible, it takes the simplified form $2w + \varepsilon n_0/p_0$ and only depends on $R$ and $\varepsilon$.
As shown in the inset to Fig. \ref{fig:2}c, the energy of the self-trapped condensate is lower than that of the ballistic condensate for small traps, where the former is stable, rendering it more energetically favourable even in the case of bistability.


\begin{figure}
    \centering
    \includegraphics[width=1.0\linewidth]{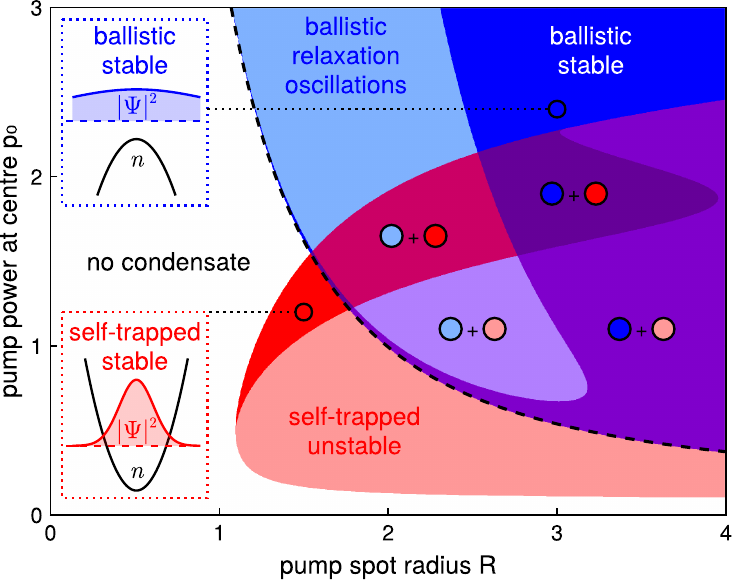}
    \caption{Numerical phase diagram of the dynamical system.
    Blue and red colors correspond to ballistic and self-trapped states, bright and light tones correspond to stable and unstable regions.
    Insets: condensate and reservoir densities $|\Psi|^2$ and $n$ computed for uniquely stable ballistic (blue) and self-trapped (red) states.
    Parameters: $\varepsilon = 8.0, \xi = 0.8, a = 0.1$.}
    \label{fig:3}
\end{figure}

Numerical analysis of the dynamical system (\ref{eq:dyn_cond},\ref{eq:dyn_res}) reveals a richer phase diagram, shown in Fig. \ref{fig:3}.
Within the existence regions of self-trapped (red) and ballistic (blue) fixed points both can be either stable (bright) or unstable (light).
Remarkably, stable self-trapped states exist for weak pump powers below the linear condensation threshold $p_0<p_\text{th}$, implying an on-off bistability region (bright red).
In turn, the ballistic fixed point can be unstable above the condensation threshold, giving rise to a relaxation-oscillation limit cycle.
This periodic cycle is characterized by slow accumulation of the reservoir density followed by abrupt energy release through stimulated scattering into the condensate (see SM).
The two stability regimes of both fixed-point types of Eqs.(\ref{eq:dyn_cond},\ref{eq:dyn_res}) can overlap, giving rise to more bistability regimes shown in Fig. \ref{fig:3} with mixed tones.

\begin{figure*}[]
    \centering
    \includegraphics[width=1.0\linewidth]{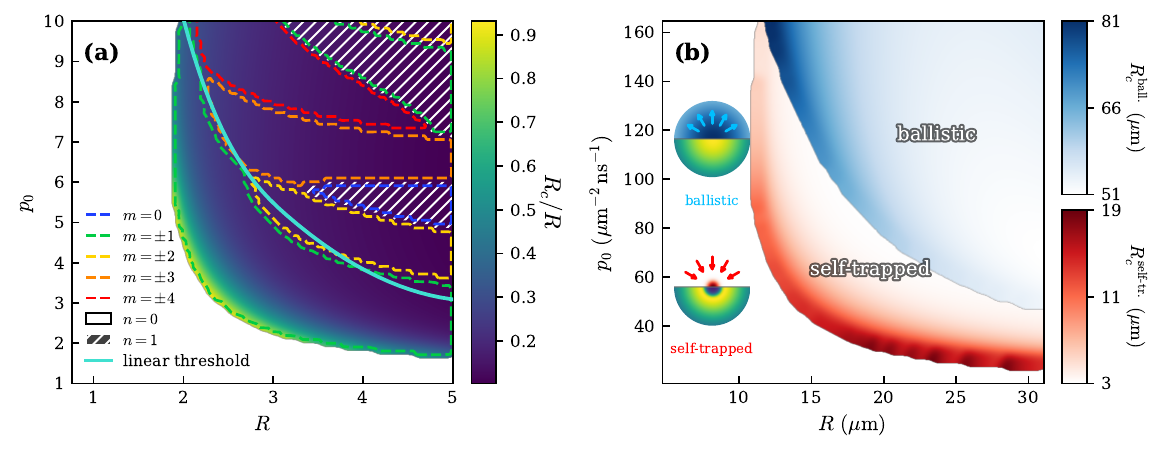}
    \caption{Numerical stability map of self-trapped and ballistic solutions obtained directly from Eqs.~\eqref{eq:eGPE}. a) $R_c/R$ of the compact self-trapped solution on the $(R,p_0)$ plane, indicating the BdG mode with the highest growth rate, i.e., the ``most unstable'', for azimuthal perturbations $m=0,\pm1,\ldots,\pm4$ (dotted lines) and the linear condensation threshold (cyan). Hatching marks the region where the dominant unstable mode has radial excitation number $n=1$ rather than $n=0$. b) Physical condensate size $R_c$ for the self-trapped (red) and ballistic (blue) solutions, converted using representative high-$Q$ microcavity parameters (see text). Insets: condensate density $|\psi|^2$ (top half) and reservoir density $\eta$ (bottom half) with arrows indicating the polariton current direction. Parameters: $\varepsilon=8, \xi=1, a=1$.}
    \label{fig:4}
\end{figure*}

Beyond the Gaussian ansatz of Eqs.~(\ref{eq:dyn_cond},\ref{eq:dyn_res}), we numerically solve the full eGPE stationary problem in polar coordinates and perform Bogoliubov-de Gennes (BdG) stability analysis for azimuthal perturbations $\delta\psi \sim e^{im\varphi}$, characterized by integer angular numbers $m$~\cite{SM}.
The family of self-trapped solutions formed by converging polariton currents consists of two branches with distinct values of condensate-to-pump-spot size ratios, in qualitative agreement with the analytic results.
In the following, we only address the compact self-trapped branch, characterized by weak instabilities as opposed to the strongly unstable wide self-trapped branch.

The existence domain of self-trapped solutions is governed by the interplay of the attractive and repulsive contributions to the effective potential.
At the critical value of the self-interaction parameter $\xi$, the two self-trapped branches annihilate at the saddle-node bifurcation (see SM for details).
While the self-trapped solutions exist both below and above the linear condensation threshold, the universally stable ballistic solution is dominant over the self-trapped solutions above the threshold.

The summary of the numerical analysis is presented in Fig. \ref{fig:4}.
Fig. ~\ref{fig:4}a. shows the domains of the dominant instability mode overlaid on the $R_c/R$ map on the $R-p_0$ plane.
Beyond the azimuthal number $m$, we further resolve the Bogoliubov excitations by their radial number $n$, corresponding to the number of antinodes in the spatial profile of the excitation mode (see \cite{SM} for details).
Fig. ~\ref{fig:4}b. shows the dimensional values of the condensate radius, as well as the linear condensation threshold separating the range of dominance of the stable ballistic state from the weakly unstable below-threshold self-trapped state.
For this illustration, we used a realistic parameter set relevant to high-$Q$ microcavities \cite{Ballarini2019}: $\Gamma^{-1}=100\,\mathrm{ps}$, $m^*=3\times10^{-4}m_e$, $\alpha_1 = 2\,\mu\mathrm{eV}\cdot\mu\mathrm{m}^2$.

In contrast to the dynamical model (\ref{eq:dyn_cond},\ref{eq:dyn_res}), the full BdG analysis reveals universal dynamical instability of the compact self-trapped solution due to excitations with positive growth rates.
These collective excitation modes are predominantly governed by the self-induced trapping potential, producing a discrete spectrum of confined modes with angular and radial numbers $m$ and $n$, respectively.
Remarkably, below the condensation threshold, the dynamical instability of the self-trapped condensate is extremely weak due to BdG excitation growth rates only marginally exceeding zero.

Numerical simulations of Eqs. \eqref{eq:eGPE} confirm weak instability of the self-trapped state, caused by growing fluctuations with nonzero angular momenta.
Once formed, the self-trapped state remains steady over extremely long times that far exceed the polariton lifetime $\Gamma^{-1}$ and even typical condensate coherence times.
However, the instability eventually leads to an abrupt ejection of the condensate from the self-induced trap in the form of a compact polariton bullet, propagating away from the pump spot in a spontaneously picked azimuthal direction (see End Matter and Supplemental Information~\cite{SM} for details). According to the provided analysis~\cite{SM}, the instability development time could reach the nanoseconds scale, which is slower than conventional condensation rate. 


In summary, we demonstrated that both self-trapped and ballistically expanding polariton condensates under CW non-resonant focused optical pumping are described as stationary solutions of the eGPE model, treating the two-component condensate-reservoir system beyond the adiabatic approximation.
Counter-intuitively, the self-trapped state becomes favourable at small pump spot sizes $R$, for which ballistic losses lead to diverging threshold pump power $p_\text{th}(R)$ \cite{Utesov2025} and render the reservoir-mediated condensate attraction dominant over standard polariton repulsion.
Most notably, the compact steady condensate residing in a self-induced potential trap, stemming from a hole burned in the reservoir density, suppresses its own ballistic losses and can thereby exist even below the linear condensation threshold pump power.

The steady self-trapped state can be directly observed under a CW nonresonant focused pump, below the formation threshold of the dynamically stable ballistically expanding condensate.
In this regime, we expect spontaneous formation of the self-trapped condensate according to two possible mechanisms.
It can be directly seeded by fluctuations of the polariton field, typically associated with a stochastic noise term in Eq. \ref{eq:eGPEcond}.
Alternatively, it can result from random fluctuations of the nonresonant pump $P$ in Eq. \eqref{eq:eGPEres}, sweeping the hysteresis loop (see Fig.\ref{fig:EM2}a) near the condensation threshold.
Regardless of the mechanism, once the self-trapped condensate is formed, it suppresses the ballistic losses by sealing itself in the gain region and can thus remain in a steady metastable state over macroscopic times until it is ejected as a compact wavepacket -- polariton bullet.

Extremely high-$Q$ microcavities sustaining long-lived polaritons, in which steady self-trapped condensates were observed \cite{Ballarini2019}, naturally favour steady condensate self-trapping.
Firstly, this is due to the longer characteristic length $L=\sqrt{\hbar/(m^*\Gamma)} \sim 10\,\mu$m, which allows reaching sufficiently low ratios $R/L\sim1$, necessary for stable self-trapping in the semi-analytic model (\ref{eq:dyn_cond},\ref{eq:dyn_res}), by using standard focused optical pumping.
Even more importantly, low damping rates $\Gamma$ of polaritonic condensate as compared to that of the excitonic reservoir $\gamma$ yield higher values of the intrinsic nonadiabaticity $a = 2\gamma/\Gamma$ and thus favour the non-adiabatic regime of self-trapping.

Both steady self-trapping and bullet ejection regimes have promising applications.
Coexistence of the self-trapped state with the absence of a condensate means an on-off bistability, which can be implemented for ultrafast all-optical binary memory.
In turn, polariton bullets can be potentially employed for synaptic connections in networks of spiking artificial neurons, providing both excitatory and inhibitory interactions between the nodes.
In general, our discovery of surprisingly rich non-adiabatic polariton condensate dynamics, induced by a single focused nonresonant beam, paves the way to studying new types of polariton condensate lattices, where synchronisation and coherence are reached in a dynamical and stochastic rather than stationary regime.

\sect{Acknowledgments}We are grateful to Hyoungsoon Choi and Yong-Hoon Cho for valuable discussions.
P.N.K. and A.V.N. acknowledge support by the Russian Science Foundation under Grant No. 25-12-00135.
O.I.U. was supported by Brain Pool Plus Program through the National Research Foundation of Korea funded by the Ministry of Science and ICT (2020H1D3A2A03099291) and National Research Foundation of Korea (NRF) grant funded by the Korea government (MSIT) (RS-2026-25470048).

\bibliography{references}

@Article{Carusotto2013,
  author    = {Carusotto, Iacopo and Ciuti, Cristiano},
  journal   = {Reviews of Modern Physics},
  title     = {Quantum fluids of light},
  year      = {2013},
  issn      = {1539-0756},
  month     = Feb,
  number    = {1},
  pages     = {299--366},
  volume    = {85},
  doi       = {10.1103/RevModPhys.85.299},
  publisher = {American Physical Society (APS)},
}

@Article{Haug2014,
  author    = {Haug, H. and Doan, T. D. and Tran Thoai, D. B.},
  journal   = {Physical Review B},
  title     = {Quantum kinetic derivation of the nonequilibrium Gross-Pitaevskii equation for nonresonant excitation of microcavity polaritons},
  year      = {2014},
  issn      = {1550-235X},
  month     = Apr,
  number    = {15},
  pages     = {155302},
  volume    = {89},
  doi       = {10.1103/PhysRevB.89.155302},
  publisher = {American Physical Society (APS)},
}

@Article{Ferrier2011,
  author    = {Ferrier, Lydie and Wertz, Esther and Johne, Robert and Solnyshkov, Dmitry D. and Senellart, Pascale and Sagnes, Isabelle and Lemaître, Aristide and Malpuech, Guillaume and Bloch, Jacqueline},
  journal   = {Physical Review Letters},
  title     = {Interactions in Confined Polariton Condensates},
  year      = {2011},
  issn      = {1079-7114},
  month     = Mar,
  number    = {12},
  pages     = {126401},
  volume    = {106},
  doi       = { 10.1103/PhysRevLett.106.126401},
  publisher = {American Physical Society (APS)},
}

@Article{Wouters2007,
  author    = {Wouters, Michiel and Carusotto, Iacopo},
  journal   = {Physical Review Letters},
  title     = {Excitations in a Nonequilibrium Bose-Einstein Condensate of Exciton Polaritons},
  year      = {2007},
  issn      = {1079-7114},
  month     = Oct,
  number    = {14},
  pages     = {140402},
  volume    = {99},
  doi       = { 10.1103/PhysRevLett.99.140402},
  publisher = {American Physical Society (APS)},
}

@Article{Schneider2017,
  author    = {Schneider, C and Winkler, K and Fraser, M D and Kamp, M and Yamamoto, Y and Ostrovskaya, E A and Höfling, S},
  journal   = {Reports on Progress in Physics},
  title     = {Exciton-polariton trapping and potential landscape engineering},
  year      = {2017},
  issn      = {1361-6633},
  month     = nov,
  number    = {1},
  pages     = {016503},
  volume    = {80},
  doi       = {10.1088/0034-4885/80/1/016503},
  publisher = {IOP Publishing},
}

@book{Kavokin2007,
  author    = {Kavokin, Alexey and Baumberg, Jeremy J. and Malpuech, Guillaume and Laussy, Fabrice P.},
  title     = {Microcavities},
  publisher = {Oxford University Press},
  address   = {Oxford; New York},
  edition   = {1st},
  year      = {2007},
  isbn      = {9780199228942},
  series    = {Series on Semiconductor Science and Technology},
  number    = {16},
}

@Article{Alyatkin2020,
  author    = {Alyatkin, S. and Töpfer, J. D. and Askitopoulos, A. and Sigurdsson, H. and Lagoudakis, P. G.},
  journal   = {Physical Review Letters},
  title     = {Optical Control of Couplings in Polariton Condensate Lattices},
  year      = {2020},
  issn      = {1079-7114},
  month     = May,
  number    = {20},
  pages     = {207402},
  volume    = {124},
  doi       = { 10.1103/PhysRevLett.124.207402},
  publisher = {American Physical Society (APS)},
}

@Article{Christmann2012,
  author    = {Christmann, Gabriel and Tosi, Guilherme and Berloff, Natalia G. and Tsotsis, Panos and Eldridge, Peter S. and Hatzopoulos, Zacharias and Savvidis, Pavlos G. and Baumberg, Jeremy J.},
  journal   = {Physical Review B},
  title     = {Polariton ring condensates and sunflower ripples in an expanding quantum liquid},
  year      = {2012},
  issn      = {1550-235X},
  month     = June,
  number    = {23},
  pages     = {235303},
  volume    = {85},
  doi       = {10.1103/PhysRevB.85.235303},
  publisher = {American Physical Society (APS)},
}

@Article{Berloff2017,
  author    = {Berloff, Natalia G. and Silva, Matteo and Kalinin, Kirill and Askitopoulos, Alexis and Topfer, Julian D. and Cilibrizzi, Pasquale and Langbein, Wolfgang and Lagoudakis, Pavlos G.},
  journal   = {Nature Materials},
  title     = {Realizing the classical XY Hamiltonian in polariton simulators},
  year      = {2017},
  issn      = {1476-4660},
  month     = Sept,
  number    = {11},
  pages     = {1120--1126},
  volume    = {16},
  doi       = {10.1038/nmat4971},
  publisher = {Springer Science and Business Media LLC},
}

@Article{Alyatkin2021,
  author    = {Alyatkin, S. and Sigurdsson, H. and Askitopoulos, A. and Töpfer, J. D. and Lagoudakis, P. G.},
  journal   = {Nature Communications},
  title     = {Quantum fluids of light in all-optical scatterer lattices},
  year      = {2021},
  issn      = {2041-1723},
  month     = Sept,
  number    = {1},
  volume    = {12},
  doi       = {10.1038/s41467-021-25845-4},
  publisher = {Springer Science and Business Media LLC},
}

@Article{Alyatkin2025,
  author    = {Alyatkin, Sergey and Sitnik, Kirill and Daníelsson, Valtýr Kári and Kartashov, Yaroslav V. and Töpfer, Julian D. and Sigurdsson, Helgi and Lagoudakis, Pavlos G.},
  journal   = {Science Advances},
  title     = {Quantum fluids of light in 2D artificial reconfigurable aperiodic crystals with tailored coupling},
  year      = {2025},
  issn      = {2375-2548},
  month     = Sept,
  number    = {39},
  volume    = {11},
  doi       = {10.1126/sciadv.adz2484},
  publisher = {American Association for the Advancement of Science (AAAS)},
}

@Article{Alyatkin2024,
  author    = {Alyatkin, Sergey and Milián, Carles and Kartashov, Yaroslav V. and Sitnik, Kirill A. and Gnusov, Ivan and Töpfer, Julian D. and Sigurdsson, Helgi and Lagoudakis, Pavlos G.},
  journal   = {Science Advances},
  title     = {Antiferromagnetic Ising model in a triangular vortex lattice of quantum fluids of light},
  year      = {2024},
  issn      = {2375-2548},
  month     = Aug,
  number    = {34},
  volume    = {10},
  doi       = {10.1126/sciadv.adj1589},
  publisher = {American Association for the Advancement of Science (AAAS)},
}

@Article{Askitopoulos2013,
  author    = {Askitopoulos, A. and Ohadi, H. and Kavokin, A. V. and Hatzopoulos, Z. and Savvidis, P. G. and Lagoudakis, P. G.},
  journal   = {Physical Review B},
  title     = {Polariton condensation in an optically induced two-dimensional potential},
  year      = {2013},
  issn      = {1550-235X},
  month     = July,
  number    = {4},
  pages     = {041308},
  volume    = {88},
  doi       = { 10.1103/PhysRevB.88.041308},
  publisher = {American Physical Society (APS)},
}

@Article{Askitopoulos2015,
  author    = {Askitopoulos, A. and Liew, T. C. H. and Ohadi, H. and Hatzopoulos, Z. and Savvidis, P. G. and Lagoudakis, P. G.},
  journal   = {Physical Review B},
  title     = {Robust platform for engineering pure-quantum-state transitions in polariton condensates},
  year      = {2015},
  issn      = {1550-235X},
  month     = jul,
  number    = {3},
  pages     = {035305},
  volume    = {92},
  doi       = {10.1103/physrevb.92.035305},
  publisher = {American Physical Society (APS)},
}

@Article{Orfanakis2021,
  author    = {Orfanakis, K. and Tzortzakakis, A. F. and Petrosyan, D. and Savvidis, P. G. and Ohadi, H.},
  journal   = {Physical Review B},
  title     = {Ultralong temporal coherence in optically trapped exciton-polariton condensates},
  year      = {2021},
  issn      = {2469-9969},
  month     = jun,
  number    = {23},
  pages     = {235313},
  volume    = {103},
  doi       = {10.1103/physrevb.103.235313},
  publisher = {American Physical Society (APS)},
}

@Article{Liu2015,
  author    = {Liu, Gangqiang and Snoke, David W. and Daley, Andrew and Pfeiffer, Loren N. and West, Ken},
  journal   = {Proceedings of the National Academy of Sciences},
  title     = {A new type of half-quantum circulation in a macroscopic polariton spinor ring condensate},
  year      = {2015},
  issn      = {1091-6490},
  month     = Feb,
  number    = {9},
  pages     = {2676--2681},
  volume    = {112},
  doi       = {10.1073/pnas.1424549112},
  publisher = {Proceedings of the National Academy of Sciences},
}

@Article{Sun2018,
  author    = {Sun, Yongbao and Yoon, Yoseob and Khan, Saeed and Ge, Li and Steger, Mark and Pfeiffer, Loren N. and West, Ken and Türeci, Hakan E. and Snoke, David W. and Nelson, Keith A.},
  journal   = {Physical Review B},
  title     = {Stable switching among high-order modes in polariton condensates},
  year      = {2018},
  issn      = {2469-9969},
  month     = Jan,
  number    = {4},
  pages     = {045303},
  volume    = {97},
  doi       = {10.1103/PhysRevB.97.045303},
  publisher = {American Physical Society (APS)},
}

@Article{Ohadi2015,
  author    = {Ohadi, H. and Dreismann, A. and Rubo, Y. G. and Pinsker, F. and del Valle-Inclan Redondo, Y. and Tsintzos, S. I. and Hatzopoulos, Z. and Savvidis, P. G. and Baumberg, J. J.},
  journal   = {Physical Review X},
  title     = {Spontaneous Spin Bifurcations and Ferromagnetic Phase Transitions in a Spinor Exciton-Polariton Condensate},
  year      = {2015},
  issn      = {2160-3308},
  month     = July,
  number    = {3},
  pages     = {031002},
  volume    = {5},
  doi       = { 10.1103/PhysRevX.5.031002},
  publisher = {American Physical Society (APS)},
}

@Article{Ohadi2017,
  author    = {Ohadi, H. and Ramsay, A.J. and Sigurdsson, H. and del Valle-Inclan Redondo, Y. and Tsintzos, S.I. and Hatzopoulos, Z. and Liew, T.C.H. and Shelykh, I.A. and Rubo, Y.G. and Savvidis, P.G. and Baumberg, J.J.},
  journal   = {Physical Review Letters},
  title     = {Spin Order and Phase Transitions in Chains of Polariton Condensates},
  year      = {2017},
  issn      = {1079-7114},
  month     = Aug,
  number    = {6},
  pages     = {067401},
  volume    = {119},
  doi       = { 10.1103/PhysRevLett.119.067401},
  publisher = {American Physical Society (APS)},
}

@Article{Dreismann2014,
  author    = {Dreismann, Alexander and Cristofolini, Peter and Balili, Ryan and Christmann, Gabriel and Pinsker, Florian and Berloff, Natasha G. and Hatzopoulos, Zacharias and Savvidis, Pavlos G. and Baumberg, Jeremy J.},
  journal   = {Proceedings of the National Academy of Sciences},
  title     = {Coupled counterrotating polariton condensates in optically defined annular potentials},
  year      = {2014},
  issn      = {1091-6490},
  month     = June,
  number    = {24},
  pages     = {8770--8775},
  volume    = {111},
  doi       = {10.1073/pnas.1401988111},
  publisher = {Proceedings of the National Academy of Sciences},
}

@Article{Pieczarka2021,
  author    = {Pieczarka, Maciej and Estrecho, Eliezer and Ghosh, Sanjib and Wurdack, Matthias and Steger, Mark and Snoke, David W. and West, Kenneth and Pfeiffer, Loren N. and Liew, Timothy C. H. and Truscott, Andrew G. and Ostrovskaya, Elena A.},
  journal   = {Optica},
  title     = {Topological phase transition in an all-optical exciton-polariton lattice},
  year      = {2021},
  issn      = {2334-2536},
  month     = Aug,
  number    = {8},
  pages     = {1084},
  volume    = {8},
  doi       = {10.1364/OPTICA.426996},
  publisher = {Optica Publishing Group},
}

@Article{Pickup2018,
  author    = {Pickup, L. and Kalinin, K. and Askitopoulos, A. and Hatzopoulos, Z. and Savvidis, P. G. and Berloff, N. G. and Lagoudakis, P. G.},
  journal   = {Physical Review Letters},
  title     = {Optical Bistability under Nonresonant Excitation in Spinor Polariton Condensates},
  year      = {2018},
  issn      = {1079-7114},
  month     = May,
  number    = {22},
  pages     = {225301},
  volume    = {120},
  doi       = {10.1103/PhysRevLett.120.225301},
  publisher = {American Physical Society (APS)},
}

@Article{Pickup2020,
  author    = {Pickup, L. and Sigurdsson, H. and Ruostekoski, J. and Lagoudakis, P. G.},
  journal   = {Nature Communications},
  title     = {Synthetic band-structure engineering in polariton crystals with non-Hermitian topological phases},
  year      = {2020},
  issn      = {2041-1723},
  month     = sep,
  number    = {1},
  volume    = {11},
  doi       = {10.1038/s41467-020-18213-1},
  publisher = {Springer Science and Business Media LLC},
}

@article{Kozhevin2025,
  author        = {P. N. Kozhevin and A. D. Liubomirov and R. V. Cherbunin and M. A. Chukeev and I. Yu. Chestnov and A. V. Kavokin and A. V. Nalitov},
  title         = {Supersolid in optically trapped exciton-polariton condensates},
  year          = {2025},
  eprint        = {2507.14585},
  journal = {arXiv},
  primaryclass  = {cond-mat.quant-gas},
  doi           = {10.48550/arXiv.2507.14585},
}

@article{Alyatkin2024APL,
  author  = {Alyatkin, Sergey and Sigurdhsson, Helgi and Kartashov, Yaroslav V. and Gnusov, Ivan and Sitnik, Kirill and T{\"o}pfer, Julian D. and Lagoudakis, Pavlos G.},
  title   = {All-optical triangular and honeycomb lattices of exciton–polaritons},
  journal = {Applied Physics Letters},
  volume  = {124},
  number  = {6},
  pages   = {062105},
  year    = {2024},
  month   = feb,
  doi     = {10.1063/5.0180272},
}

@Article{Cookson2021,
  author    = {Cookson, Tamsin and Kalinin, Kirill and Sigurdsson, Helgi and Topfer, Julian D. and Alyatkin, Sergey and Silva, Matteo and Langbein, Wolfgang and Berloff, Natalia G. and Lagoudakis, Pavlos G.},
  journal   = {Nature Communications},
  title     = {Geometric frustration in polygons of polariton condensates creating vortices of varying topological charge},
  year      = {2021},
  issn      = {2041-1723},
  month     = apr,
  number    = {1},
  volume    = {12},
  doi       = {10.1038/s41467-021-22121-3},
  publisher = {Springer Science and Business Media LLC},
}

@Article{Ballarini2019,
  author    = {Ballarini, Dario and Chestnov, Igor and Caputo, Davide and De Giorgi, Milena and Dominici, Lorenzo and West, Kenneth and Pfeiffer, Loren N. and Gigli, Giuseppe and Kavokin, Alexey and Sanvitto, Daniele},
  journal   = {Physical Review Letters},
  title     = {Self-Trapping of Exciton-Polariton Condensates in GaAs Microcavities},
  year      = {2019},
  issn      = {1079-7114},
  month     = July,
  number    = {4},
  pages     = {047401},
  volume    = {123},
  doi       = { 10.1103/PhysRevLett.123.047401},
  publisher = {American Physical Society (APS)},
}

@Article{Dominici2015,
  author    = {Dominici, L. and Petrov, M. and Matuszewski, M. and Ballarini, D. and De Giorgi, M. and Colas, D. and Cancellieri, E. and Silva Fernandez, B. and Bramati, A. and Gigli, G. and Kavokin, A. and Laussy, F. and Sanvitto, D.},
  journal   = {Nature Communications},
  title     = {Real-space collapse of a polariton condensate},
  year      = {2015},
  issn      = {2041-1723},
  month     = Dec,
  number    = {1},
  volume    = {6},
  doi       = {10.1038/ncomms9993},
  publisher = {Springer Science and Business Media LLC},
}

@Article{Estrecho2018,
  author    = {Estrecho, E. and Gao, T. and Bobrovska, N. and Fraser, M. D. and Steger, M. and Pfeiffer, L. and West, K. and Liew, T. C. H. and Matuszewski, M. and Snoke, D. W. and Truscott, A. G. and Ostrovskaya, E. A.},
  journal   = {Nature Communications},
  title     = {Single-shot condensation of exciton polaritons and the hole burning effect},
  year      = {2018},
  issn      = {2041-1723},
  month     = aug,
  number    = {1},
  volume    = {9},
  doi       = {10.1038/s41467-018-05349-4},
  publisher = {Springer Science and Business Media LLC},
}

@article{Bobrovska2018,
  author    = {Bobrovska, Nataliya and Matuszewski, Michal and Daskalakis, Konstantinos S. and Maier, Stefan A. and Kéna-Cohen, Stéphane},
  title     = {Dynamical Instability of a Nonequilibrium Exciton-Polariton Condensate},
  journal   = {ACS Photonics},
  volume    = {5},
  number    = {1},
  pages     = {111--118},
  year      = {2018},
  doi       = {10.1021/acsphotonics.7b00283},
}

@Article{Ohadi2016,
  author    = {Ohadi, H. and Gregory, R.L. and Freegarde, T. and Rubo, Y.G. and Kavokin, A.V. and Berloff, N.G. and Lagoudakis, P.G.},
  journal   = {Physical Review X},
  title     = {Nontrivial Phase Coupling in Polariton Multiplets},
  year      = {2016},
  issn      = {2160-3308},
  month     = Aug,
  number    = {3},
  pages     = {031032},
  volume    = {6},
  doi       = { 10.1103/PhysRevX.6.031032},
  publisher = {American Physical Society (APS)},
}

@Article{Sigurdsson2019,
  author    = {Sigurdsson, H. and Krivosenko, Y. S. and Iorsh, I. V. and Shelykh, I. A. and Nalitov, A. V.},
  journal   = {Physical Review B},
  title     = {Spontaneous topological transitions in a honeycomb lattice of exciton-polariton condensates due to spin bifurcations},
  year      = {2019},
  issn      = {2469-9969},
  month     = Dec,
  number    = {23},
  pages     = {235444},
  volume    = {100},
  doi       = { 10.1103/PhysRevB.100.235444},
  publisher = {American Physical Society (APS)},
}

@Article{Utesov2025,
  author    = {Utesov, Oleg I. and Park, Min and Choi, Daegwang and Choi, Soohong and Park, Suk In and Kang, Sooseok and Song, Jin Dong and Osipov, Alexey N. and Yulin, Alexey V. and Cho, Yong-Hoon and Choi, Hyoungsoon and Aranson, Igor S. and Koniakhin, Sergei V.},
  journal   = {Communications Physics},
  title     = {Universal condensation threshold dependence on pump beam size for exciton-polaritons},
  year      = {2025},
  issn      = {2399-3650},
  month     = July,
  number    = {1},
  volume    = {8},
  doi       = {10.1038/s42005-025-02198-8},
  publisher = {Springer Science and Business Media LLC},
}

@Article{Nalitov2019,
  author    = {Nalitov, A. V. and Sigurdsson, H. and Morina, S. and Krivosenko, Y. S. and Iorsh, I. V. and Rubo, Y. G. and Kavokin, A. V. and Shelykh, I. A.},
  journal   = {Physical Review A},
  title     = {Optically trapped polariton condensates as semiclassical time crystals},
  year      = {2019},
  issn      = {2469-9934},
  month     = Mar,
  number    = {3},
  pages     = {033830},
  volume    = {99},
  doi       = {10.1103/physreva.99.033830},
  publisher = {American Physical Society (APS)},
}

@Article{Balas2026,
  author    = {Balas, Yannis C. and Zhou, Xiaoqing and Cherotchenko, Evgeniia and Kuznetsov, Ivan and Rajendran, Sai Kiran and Paschos, Giannis G. and Trifonov, Artur V. and Nalitov, Anton and Ohadi, Hamid and Savvidis, Pavlos G.},
  journal   = {Science Bulletin},
  title     = {Ultra-small mode volume polariton condensation via precision He+ implantation},
  year      = {2026},
  issn      = {2095-9273},
  month     = July,
  number    = {14},
  pages     = {3617--3623},
  volume    = {71},
  doi       = {10.1016/j.scib.2026.06.026},
  publisher = {Elsevier BV},
}

@Article{Barrat2024,
  author    = {Barrat, Joris and Tzortzakakis, Andreas F. and Niu, Meng and Zhou, Xiaoqing and Paschos, Giannis G. and Petrosyan, David and Savvidis, Pavlos G.},
  journal   = {Science Advances},
  title     = {Qubit analog with polariton superfluid in an annular trap},
  year      = {2024},
  issn      = {2375-2548},
  month     = Oct,
  number    = {43},
  volume    = {10},
  doi       = {10.1126/sciadv.ado4042},
  publisher = {American Association for the Advancement of Science (AAAS)},
}

@Article{Chiao1964,
  author    = {Chiao, R. Y. and Garmire, E. and Townes, C. H.},
  journal   = {Physical Review Letters},
  title     = {Self-Trapping of Optical Beams},
  year      = {1964},
  issn      = {0031-9007},
  month     = Oct,
  number    = {15},
  pages     = {479--482},
  volume    = {13},
  doi       = {10.1103/physrevlett.13.479},
  publisher = {American Physical Society (APS)},
}

@Article{Ostrovskaya2012,
  author    = {Ostrovskaya, Elena A. and Abdullaev, Jasur and Desyatnikov, Anton S. and Fraser, Michael D. and Kivshar, Yuri S.},
  journal   = {Physical Review A},
  title     = {Dissipative solitons and vortices in polariton Bose-Einstein condensates},
  year      = {2012},
  issn      = {1094-1622},
  month     = July,
  number    = {1},
  pages     = {013636},
  volume    = {86},
  doi       = {10.1103/physreva.86.013636},
  publisher = {American Physical Society (APS)},
}

@misc{SM,
  note = {See Supplemental Material at [URL to be inserted by publisher] for details of numerical calculations and experimental sample.}
}

@misc{animation,
  note = {See accompanying animation at [URL to be inserted by publisher]}
}

@Article{Hu2026,
  author    = {Hu, Junwei and Idrees, Muhammad and Zhang, Kun and Lin, Ji and Li, Hui-jun and Kavokin, Alexey},
  journal   = {Physical Review B},
  title     = {Self-localized solitonlike peak in uniform nonresonantly pumped exciton-polariton condensates},
  year      = {2026},
  issn      = {2469-9969},
  month     = Feb,
  number    = {7},
  volume    = {113},
  doi       = {10.1103/kw8l-mh6q},
  publisher = {American Physical Society (APS)},
}

@article{Guizar2004,
  author  = {Guizar-Sicairos, Manuel and Guti\'errez-Vega, Julio C.},
  title   = {Computation of quasi-discrete Hankel transforms of integer order for propagating optical wave fields},
  journal = {Journal of the Optical Society of America A},
  volume  = {21},
  number  = {1},
  pages   = {53--58},
  year    = {2004},
  doi     = {10.1364/JOSAA.21.000053}
}

\clearpage
\setcounter{figure}{0}
\setcounter{equation}{0}
\renewcommand{\thefigure}{A\arabic{figure}}


\appendix*

\section{End Matter}

\subsection{Weakly nonlinear regime}

\begin{figure}[h!]
    \centering
    \includegraphics[width=1.0\linewidth]{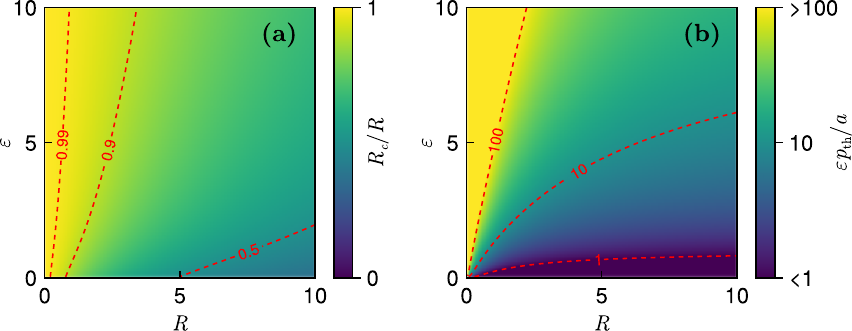}
    \caption{Weakly nonlinear regime analysis. a) The universal ratio of the condensate and pump spot sizes $R_c/R<1$. b) The effective reservoir-mediated attraction parameter $A \equiv \varepsilon p_\text{th}/a$.}
    \label{fig:EM1}
\end{figure}

The radius $R_c(\varepsilon,R)$ of a two-dimensional condensate at the linear threshold of Gaussian incoherent pumping is given by the explicit universal law \cite{Utesov2025}:
\begin{equation}
    R_c = \mathrm{Re} \left\{ - {i \over R} \sqrt{\varepsilon + i} \left(I(\varepsilon,R) + \sqrt{I(\varepsilon,R)^2 + 1}\right) \right\}^{-1/2}, \nonumber
\end{equation}
where $I(\varepsilon,R) = \mathrm{Im \{ {i} \sqrt{(\varepsilon+i)}\}}/R$.
The ratio of the near-threshold condensate and pump spot sizes $R_c/R<1$ is shown in Fig. \ref{fig:EM1}a.

The peak strength of the effective reservoir-mediated attraction at the linear condensation threshold is conveniently illustrated on the same parameter plane, with the value $A(\varepsilon,R)$ given by
\begin{equation}
    A \equiv \varepsilon p_\text{th}/a = \varepsilon \left( I(\varepsilon,R) + \sqrt{I(\varepsilon,R)^2+1} \right)^2.
\end{equation}
The condition of effective attraction dominance $\xi^\prime > \xi$ is recast in terms of $A(\varepsilon,R)$ as $A>\xi a$.
Recent estimates for the interaction parameter $\varepsilon\sim1$ \cite{Balas2026} combined with observations of oscillating polariton vortex \cite{Barrat2024} indicating $\xi\sim\varepsilon$ \cite{Nalitov2019}, give a rough estimate $\xi\sim1$ for the first multiplier.
As the lifetime of the predominantly excitonic reservoir typically exceeds that of the more photonic condensate, the second multiplier is limited by $a<1$.
One thus concludes that the reservoir-mediated attraction mechanism is dominant near the condensation threshold in a wide range of parameters for which $A\gtrsim1$, as shown in Fig. \ref{fig:EM1}b.


\subsection{Deeply nonlinear regime}

In the limiting case $|\psi|^2\gg a$, the reservoir density takes the simple form:
\begin{equation}
    \eta \approx {p \over |\psi|^2} = {p_0 \over n_0} \exp\left[ {\rho^2 \left( {1 \over R_c^2} - {1 \over R^2}\right) } \right],
\end{equation}
corresponding to the leading-order radial components
\begin{equation}
    \eta_0 = {p_0 \over n_0}, \quad \eta_2 = {p_0 \over n_0} \left( {1 \over R_c^2} - {1 \over R^2} \right) 
\end{equation}
This introduces the effective non-Hermitian potential:
\begin{equation}
    u(\rho) \approx (\varepsilon + i) (\eta_0 + \eta_2 \rho^2).
\end{equation}

Self-consistency requires that the condensate wavefunction is an eigenstate of the non-Hermitian self-induced potential yielding the condition for $s$:
\begin{equation} \label{eq:s}
    s^2 = \left(\varepsilon + i\right) {p_0 \over n_0} \left( \mathrm{Re}\{ s \} - {1 \over R^2} \right).
\end{equation}
Two self-trapped solutions ($R_c<R$) exist for Eq. \eqref{eq:s}:
\begin{align}
    s_\pm =& {p_0 \over 4 n_0} \left[ 1 \pm \sqrt{1 - {8  n_0 \over p_0 R^2 (\varepsilon+\sqrt{\varepsilon^2+1})}} \right] \times \nonumber \\
    &\left( \varepsilon + \sqrt{\varepsilon^2+1} +i\right),
\end{align}
implicitly allowed for sufficiently low condensate density
\begin{equation}
    {n_0 \over p_0} < { R^2 \over 8 } \left( \varepsilon+\sqrt{\varepsilon^2 + 1} \right).
\end{equation}
The density, in turn, follows from the gain-loss balance:
\begin{equation} \label{eq:stable_condensate_density}
    {n_0 \over p_0} = t \pm \sqrt{t^2-2}, \quad t \equiv {3 \over 2} - {1 \over R^2 (\varepsilon + \sqrt{\varepsilon^2 + 1})}.
\end{equation}
Both solutions exist for sufficiently large pumping spots:
\begin{equation}
    R>R_\text{crit} = \left[ \left( \varepsilon + \sqrt{\varepsilon^2 + 1} \right) \left({3 \over 2} - \sqrt{2} \right)\right]^{-{1/2}}.
\end{equation}
In contrast, only one ballistic solution with $R_c>R$ exists for any values of $\varepsilon$, $R$:
\begin{align}
    s =& {p_0 \over 4 |\psi_0|^2} \left[ \sqrt{1 + {8 n_0 \over p_0 R^2 (\sqrt{\varepsilon^2+1}-\varepsilon)}} - 1\right] \times \nonumber \\
    &\left( \sqrt{\varepsilon^2+1} - \varepsilon -i\right),
\end{align}
with condensate density given by
\begin{equation}
     {n_0 \over p_0} = t - \sqrt{t^2-2}, \quad t \equiv {3 \over 2} + {1 \over R^2 (\sqrt{\varepsilon^2 + 1}-\varepsilon)}.
\end{equation}


\subsection{Dynamical instabilities}

The stability of the fixed points of the dynamical system (\ref{eq:dyn_cond}-\ref{eq:dyn_res}) is governed by the spectrum of the Jacobi matrix
    \begin{equation}
        \left| \begin{matrix}
            2(\eta_0 - 1 - 2b) & 0 & -4n_0 & 2n_0 & 0 \\
            0 & -4b & -4w & 0 & -2 \\
            2\xi w & 4w + 2 \xi n_0 & -4b & 0 & -2\varepsilon \\
            -\eta_0 & 0 & 0 & -(a+n_0) & 0 \\
            w\eta_0 - \eta_2 & n_0 \eta_0 & 0 & n_0 w & -(a + n_0)
        \end{matrix} \right|.
    \end{equation}

\begin{figure}[]
    \centering
    \includegraphics[width=1.0\linewidth]{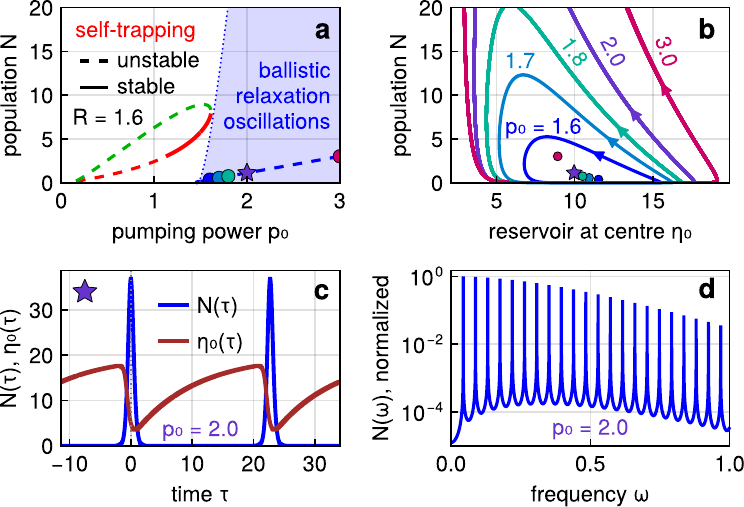}
    \caption{Relaxation oscillation regime of the system (\ref{eq:dyn_cond},\ref{eq:dyn_res}).
    a) Condensate population $N$ as a function of peak pump power $p_0$ for stationary self-trapped (red and green) and limit-cycle ballistic (blue) states.
    The stable self-trapped state (solid red) is present below the linear condensation threshold $p_\text{th}\approx1.5$.
    Unstable ballistic fixed points (dashed blue) give rise to the relaxation-oscillation limit cycle.
    b) Phase-space trajectories, condensate population $N$ vs the reservoir density $\eta_0$, for different pump powers $p_0$. Unstable fixed points are shown as dots of respective colors.
    c) Periodic time dependence of $N$ and $\eta_0$ for $p_0=2.0$. 
    d) Frequency comb in the Fourier spectrum of the condensate population $N(\omega)$.
    }
    \label{fig:EM2}
\end{figure}

Instability of the self-trapped solution below the linear condensation threshold (pink region in Fig. \ref{fig:3}) leaves the trivial solution with no condensate uniquely stable, in contrast to the on-off bistability case.
Instability of the ballistic solution, which only exists above the linear condensation threshold, gives rise to relaxation-oscillation limit cycle, illustrated in Fig. \ref{fig:EM2}.
This regime is conveniently illustrated on the phase plane of the condensate population $N = \int_0^{2\pi} d \varphi \int_0^\infty |\psi(\rho)|^2 \rho d\rho = \pi n_0/w$ and the reservoir density $\eta_0$.
The former exhibits sharp energy-releasing peaks of height increasing with the pump power (blue dotted line in Fig. \ref{fig:EM2}), while the latter slowly accumulates at the quasi-linear cycle stage between peaks, where the condensate is virtually absent, as shown in Figs.~\ref{fig:EM2}b,c.
This dynamics leads to a frequency comb in the Fourier spectrum of the condensate emission intensity, proportional to its total population, as shown in Fig.~\ref{fig:EM2}d.

\begin{figure}[b]
    \centering
    \includegraphics[width=1.0\linewidth]{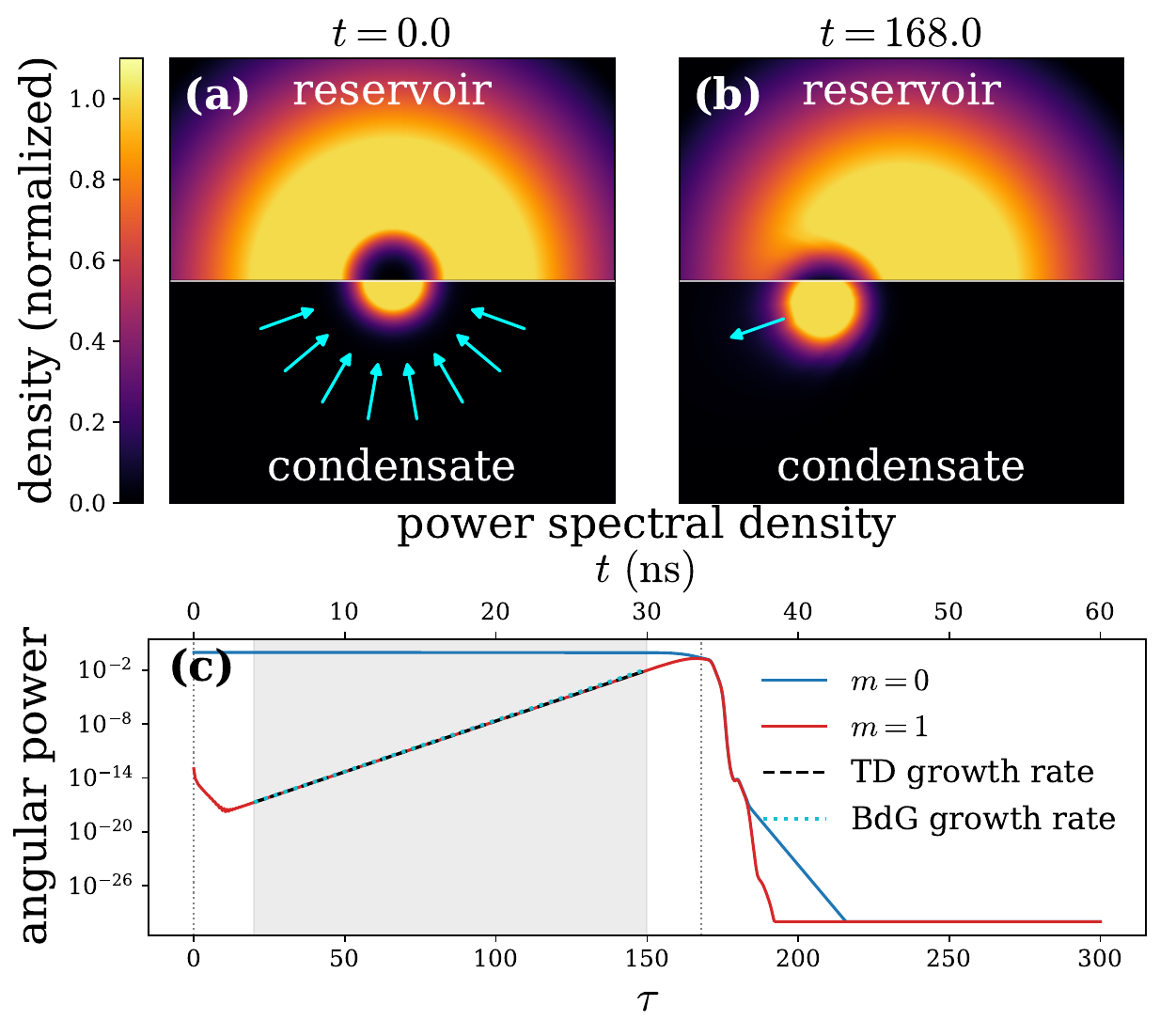}
    \caption{Weak $m=1$ instability of the self-trapped state below the linear threshold.
    a,b) Reservoir (top half) and condensate (bottom half) density at $t=0$ and $t=168$; arrows indicate the polariton current direction.
    c) Power spectral density of the $m=0$ and $m=1$ angular harmonics of the condensate; the time-domain (TD) growth rate agrees with the BdG prediction. $\Gamma^{-1}=100\,\mathrm{ps}$ was used to restore the dimension of time.}
    \label{fig:EM3}
\end{figure}


\subsection{Weak instability of the self-trapped state}

Stationary solutions of Eqs.~\eqref{eq:eGPE} support a marginally unstable compact self-trapped state below the linear condensation threshold, where the ballistic fixed point does not exist and the self-trapped condensate family coexists only with the trivial solution.
Fig. \ref{fig:EM3}. illustrates this weak instability: the stationary self-trapped solution is used as the initial condition for time-dependent integration in polar coordinates~\cite{SM}, seeded with a small white-noise $m=1$ perturbation.
Fig.~\ref{fig:EM3}a. shows the reservoir (top) and condensate (bottom) density at $t=0$, with arrows schematically indicating the particle current.
At $t=168$ (Fig. \ref{fig:EM3}b.), the $m=1$, $n=0$ BdG mode breaks the angular symmetry and ejects a polariton bullet.
Fig. \ref{fig:EM3}c. shows the power spectral density of the $m=0$ and $m=1$ angular harmonics; the growth rate extracted directly from the time-dependent density (black dashed) matches the BdG prediction (cyan dotted).
Following the bullet-ejection transient, the system relaxes into the trivial state.
The direction of bullet movement is defined by a chosen perturbation noise seed (spontaneous in nature) only in the absence of external symmetry breaking.
Cartesian coordinate simulations show the pinning of bullet orientation along simulation domain diagonals due to square grid symmetry, which is why implementing numerical methods based on polar coordinates was essential.
For a high-$Q$ microcavity with polariton lifetime $\Gamma^{-1}=100\,\mathrm{ps}$, the corresponding growth rate $\mathrm{Im}(\omega)=0.1314$ translates into a characteristic growth time of $\approx 1.5\,\mathrm{ns}$, opening a realistic window for experimental observation of the self-trapped condensate below threshold.
The time dynamics is illustrated in the accompanying animation~\cite{animation}.

\clearpage
\setcounter{equation}{0}
\setcounter{figure}{0}
\renewcommand{\theequation}{S.\arabic{equation}}
\renewcommand{\thefigure}{S\arabic{figure}}

\section*{supplementary materials}
\section{Numerical methods}

\subsection{Dimensionless dynamical equations}

The dimensionless dynamical equations (see Main Text) read
\begin{align}
    i\partial_\tau\psi &= \left[-\nabla_\rho^2 + \varepsilon\eta + \xi|\psi|^2 - i(\eta-1)\right]\psi, \label{eq:sm_dyn_psi} \\
    \partial_\tau\eta &= p - (a+|\psi|^2)\eta, \label{eq:sm_dyn_eta}
\end{align}
with $p(\rho)=p_0\exp(-\rho^2/R^2)$.

\subsection{Dimensionless stationary equations}

Stationary solutions $\psi(\rho,\theta,\tau)=e^{-i\mu\tau}\phi(\rho,\theta)$ of Eqs.~\eqref{eq:sm_dyn_psi},\eqref{eq:sm_dyn_eta} satisfy
\begin{align}
    0 &= \left[-\nabla_\rho^2 + \varepsilon\eta + \xi|\phi|^2 - \mu + i(\eta-1)\right]\phi, \label{eq:sm_stat} \\
    \eta(\rho,\theta) &= \frac{p(\rho)}{a+|\phi|^2}, \label{eq:sm_res}
\end{align}
where Eq.~\eqref{eq:sm_res} follows exactly (not only adiabatically) from $\partial_\tau\eta=0$ in Eq.~\eqref{eq:sm_dyn_eta}.

\subsection{Quasi-discrete Hankel transform (QDHT)}

For an order-$m$ angular sector, we use the quasi-discrete Hankel transform of Guizar-Sicairos and Guti\'errez-Vega \cite{Guizar2004}. Let $\alpha_1^{(m)}<\dots<\alpha_{N_r}^{(m)}<\alpha_{N_r+1}^{(m)}\equiv S^{(m)}$ be the first $N_r+1$ zeros of $J_m$. Collocation radii and conjugate wavenumbers are
\begin{equation}
    r_i^{(m)} = \alpha_i^{(m)} \frac{\rho_{\max}}{S^{(m)}}, \qquad k_i^{(m)} = \frac{\alpha_i^{(m)}}{\rho_{\max}},
\end{equation}
with transform matrix $M_{ij}^{(m)}=J_m(\alpha_i^{(m)}\alpha_j^{(m)}/S^{(m)})$. Since $J_m(kr)$ diagonalizes the order-$m$ radial Laplacian, the discretized operator
\begin{equation}
    K^{(m)} = M^{(m)}\,\mathrm{diag}\!\left[(k_i^{(m)})^2\right]\,(M^{(m)})^{-1}
\end{equation}
approximates $-\nabla_\rho^2$ acting on order-$m$ fields. For $m=0$ the radial derivative operator is built from $\partial_r J_0(kr)=-kJ_1(kr)$.

\subsection{Fixed points ($m=0$)}

Stationary axisymmetric solutions of Eq.~\eqref{eq:sm_stat} are found on the $m=0$ QDHT grid. Unknowns $x=(\mathrm{Re}\,\phi_i,\mathrm{Im}\,\phi_i,\mu)$ are fixed by the gauge $\mathrm{Im}\,\phi(r_0)=0$ ($r_0$ is innermost collocation node) and solved via Powell's hybrid Newton method with the exact analytic Jacobian of Eq.~\eqref{eq:sm_stat}. To suppress reflection of outgoing polariton currents at the truncated domain boundary $\rho_{\max}$, an absorbing potential is added, $\eta-1\to\eta-1-V_{\rm abs}(\rho)$, with $V_{\rm abs}(\rho)=0$ for $\rho<\rho_{\rm abs}$ and $V_{\rm abs}(\rho)=V_0\left(\frac{\rho-\rho_{\rm abs}}{\rho_{\max}-\rho_{\rm abs}}\right)^{n_{\rm abs}}$ otherwise.

\subsection{Time dynamics}

The full field $\psi(\rho,\theta,\tau)$ is evolved on a mixed grid: $N_r$ physical radii (the $m=0$ QDHT nodes) and $N_\theta$ uniform angular nodes. Each timestep performs a symmetric (Strang) split of Eqs.~\eqref{eq:sm_dyn_psi},\eqref{eq:sm_dyn_eta}: a half-step reservoir update, a half-kinetic step, a half-absorbing step, an exact local nonlinear kick, a half-absorbing step, a half-kinetic step, and a final half-step reservoir update. The local substep is exact at fixed $\rho,\theta$:
\begin{align}
    \eta(\tau+h) &= \eta_\infty + [\eta(\tau)-\eta_\infty]e^{-ch}, \quad \eta_\infty=\frac{p}{c},\ c=a+|\psi|^2, \\
    \psi(\tau+h) &= \psi(\tau)\,e^{(\eta-1)h}\,e^{-i(\xi|\psi|^2+\varepsilon\eta)h}.
\end{align}
The kinetic half-step is spectral: an FFT in $\theta$ yields angular components $\psi_m(\rho)$ for $|m|\leq m_{\rm considered}$ (higher $|m|$s are truncated); restricting to $m_{\rm considered}$ gives a substantial speedup, as the relevant instability physics is carried by the first few angular harmonics. Each $\psi_m$ is cubic-spline interpolated onto its own natural order-$|m|$ QDHT grid $r_i^{(m)}$, propagated by the exact free-kinetic phase $\exp[i\lambda_i^{(m)}\Delta\tau_{\rm eff}]$ with $\lambda_i^{(m)}=-(k_i^{(m)})^2$, transformed back, and interpolated onto the physical grid; an inverse FFT in $\theta$ completes the step.

\subsection{Bogoliubov-de Gennes stability}

Perturbations of an axisymmetric background $(\phi_0(\rho),\eta_0(\rho),\mu)$ are parametrized as \cite{Wouters2007}
\begin{align}
    \psi &= e^{-i\mu\tau}\left[\phi_0(\rho) + u(\rho)e^{i(m\theta-\omega\tau)} + v^*(\rho)e^{-i(m\theta-\omega^*\tau)}\right], \\
    \eta &= \eta_0(\rho) + \delta\eta(\rho)e^{i(m\theta-\omega\tau)} + \mathrm{c.c.}
\end{align}
Since $\phi_0,\eta_0$ depend on $\rho$ only, linearizing Eqs.~\eqref{eq:sm_dyn_psi},\eqref{eq:sm_dyn_eta} to first order in $(u,v,\delta\eta)$ does not couple different $m$: Fourier harmonics $e^{im\theta}$ diagonalize the axisymmetric background operators exactly, so each $|m|$ can be evaluated independently on its own order-$|m|$ QDHT grid, reducing the 2D stability problem to a sequence of 1D radial eigenproblems
\begin{equation}
    H^{(m)}\begin{pmatrix}u\\v\\\delta\eta\end{pmatrix} = \omega\begin{pmatrix}u\\v\\\delta\eta\end{pmatrix}, \quad
    H^{(m)} = \begin{pmatrix} A & B & C \\ -B^* & -A^* & -C^* \\ R_u & R_v & R_\eta \end{pmatrix},
\end{equation}
with (all diagonal in the QDHT basis, $K^{(m)}$ the order-$|m|$ Laplacian)
\begin{align}
    A &= K^{(m)} + \mathrm{diag}\!\left[\varepsilon\eta_0+2\xi|\phi_0|^2-\mu+i(\eta_0-1-V_{\rm abs})\right], \nonumber\\
    B &= \mathrm{diag}(\xi\phi_0^2), \quad C = \mathrm{diag}[(\varepsilon+i)\phi_0], \nonumber \\
    R_u &= -i\,\mathrm{diag}(\eta_0\phi_0^*), \quad R_v = -i\,\mathrm{diag}(\eta_0\phi_0), \quad R_\eta = -i\,\mathrm{diag}(a+|\phi_0|^2). \nonumber
\end{align}
Stability requires $\mathrm{Im}(\omega)\leq0$ for all resolved modes. Finite QDHT truncation introduces spurious high-$k$ eigenvalues; the leading physical mode is selected as the one with the largest $\mathrm{Im}(\omega)$ among eigenvalues satisfying $|\mathrm{Re}(\omega)|<\tfrac{2}{3}\max_i(k_i^{(m)})^2$.
\subsection{Numerical grid}
The results shown in this Supplemental Material and in the main text were obtained with $N_r=1024$ radial QDHT nodes, $\rho_{\max}=64$, $m_{\rm considered}=8$, and the timestep $\Delta\tau=0.005$ for the time-dependent runs. The absorbing potential is parametrized by $\rho_{\rm abs}=0.7\,\rho_{\max}$, $V_0=10$, and $n_{\rm abs}=3$. Convergence with respect to $N_r$ and $\Delta\tau$ was verified by repeating representative runs at increased resolution and reduced timestep and confirming that all reported quantities (fixed-point observables, BdG growth rates, and time-domain dynamics) remained unchanged within numerical tolerance.

\section{Additional results}

\subsection{Self-trapped solutions family}

The self-trapped fixed points of Eq.~\eqref{eq:sm_stat} form two branches, wide and compact, arising from a competition between the reservoir-mediated attraction and the repulsive self-interaction $\xi$. This attraction is made explicit by adiabatically eliminating the reservoir at low condensate density: expanding $\eta=p/(a+|\phi|^2)$ to first order in $|\phi|^2$,
\begin{equation}
    \eta \approx \frac{p}{a} - \frac{p}{a^2}|\phi|^2,
\end{equation}
and substituting into the potential term of Eq.~\eqref{eq:sm_stat} gives
\begin{equation}
    \varepsilon\eta + \xi|\phi|^2 \approx \frac{\varepsilon p}{a} + \left(\xi - \frac{\varepsilon p}{a^2}\right)|\phi|^2,
\end{equation}
i.e., an effective self-interaction $\xi_{\rm eff}=\xi-\varepsilon p/a^2$, which becomes attractive once $\varepsilon p/a^2>\xi$. This attraction is opposed by the bare repulsive self-interaction $\xi|\phi|^2$; increasing $\xi$ progressively weakens the net attractive potential.

\begin{figure}[]
    \centering
    \includegraphics[width=1.0\linewidth]{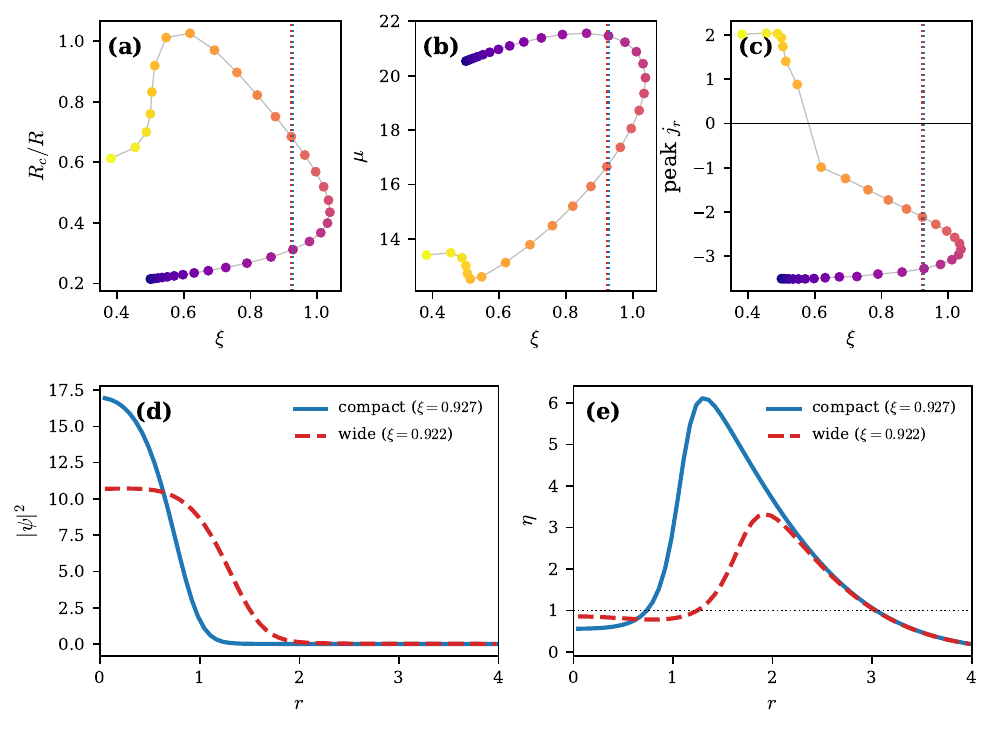}
    \caption{Self-trapped solutions family. (a) Relative size $R_c/R$, (b) condensate energy $\mu$, and (c) peak radial current $j_r$ of the wide and compact branches as functions of $\xi$ (examples of each state are marked by the vertical dotted line); the two branches annihilate in a saddle-node bifurcation at the critical value. (d,e) Condensate density $|\psi|^2$ and reservoir density $\eta$ radial profiles for the compact and wide solutions just below the bifurcation. Parameters: $\varepsilon=8,a=1, R = 2, P_0 = 10$.}
    \label{fig:S1}
\end{figure}

Figure~\ref{fig:S1} shows how increasing $\xi$ overcomes the attraction: the wide and compact branches approach each other and annihilate in a saddle-node bifurcation at a critical $\xi_c$, which sets the existence boundary of the self-trapped solution family. For sufficiently small $\xi$, an additional fixed point with radial excitation number $n=1$ appears (positive peak current, $R_c/R<1$), coexisting with the conventional $n=0$ compact solution.

\begin{figure}[]
    \centering
    \includegraphics[width=1.0\linewidth]{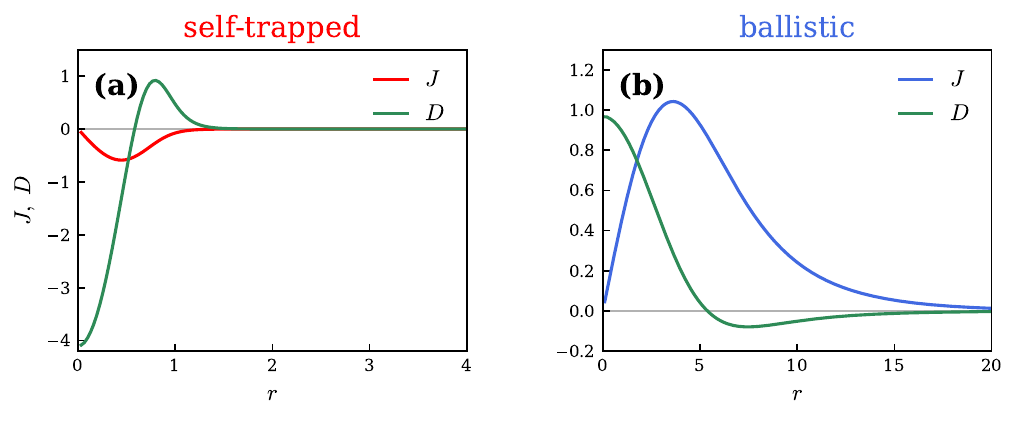}
    \caption{Radial polariton current $J(r)$ and its divergence $D(r)=\nabla\cdot\vec J$ (green) for (a) the compact self-trapped and (b) the ballistic solution of Eq.~\eqref{eq:sm_stat}, at the same pump parameters $R=5$, $p_0\approx4$. The self-trapped state exhibits a sink at the core ($D<0$) fed by a source at the boundary ($D>0$), corresponding to centripetal current $J<0$; the ballistic state exhibits the opposite topology, a source at the core and a sink on the wings, with purely outward current $J>0$. Parameters: $\varepsilon=8,\xi=1,a=1$.}
    \label{fig:S2}
\end{figure}

The main distinction between the self-trapped and ballistic condensate states can be illustrated by the behavior of the particle flux $J(r)$ (Figure~\ref{fig:S2}). In the ballistic case, $J(r)$ is non-negative everywhere, whereas for the self-trapped solution $J(r)$ shows sharply the region of opposite sign, with negative (inward current) values at the core.

\subsection{Excitation spectra structure}

\begin{figure}[]
    \centering
    \includegraphics[width=1.0\linewidth]{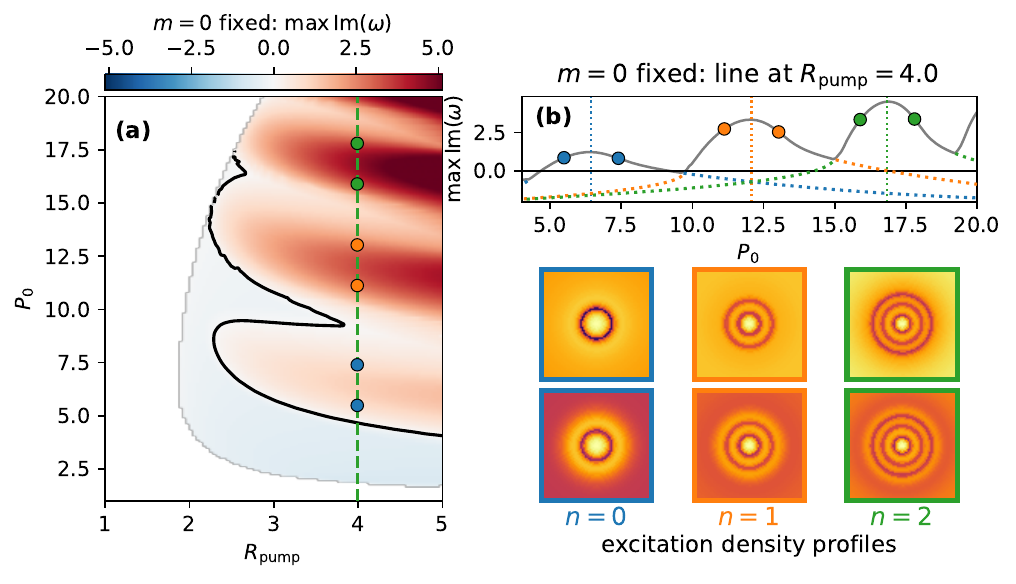}
    \caption{$m=0$ excitations spectra structure at fixed $R_{\rm pump}=4$. (a) $\max\,\mathrm{Im}(\omega)$ on the $(R_{\rm pump},P_0)$ plane. (b) $\max\,\mathrm{Im}(\omega)$ along the dashed line in (a), decomposed into contributions from successive radial excitation branches $n=0,1,2$ (colored dotted curves); mode competition between branches produces a periodic envelope as $P_0$ increases. Markers indicate the points shown below. Bottom: excitation density profiles for $n=0,1,2$ at the marked points, illustrating the increasing number of radial rings with $n$. Parameters: $\varepsilon=8,\xi=1,a=1$.}
    \label{fig:S3}
\end{figure}

The radial ($m=0$) Bogoliubov excitation spectrum of the compact self-trapped solution consists of a discrete set of branches indexed by the radial number $n$, defined as the number of rings in the excitation's spatial density profile. As the pump power $P_0$ is increased at fixed pump spot radius, successive branches compete for dominance: the observed leading instability at each $P_0$ is the mode with the largest $\mathrm{Im}(\omega)$, so the dominant $n$ switches whenever a higher branch overtakes the currently leading one.

\subsection*{Weak instability below threshold}

\begin{figure}[]
\centering
\includegraphics[width=0.8\linewidth]{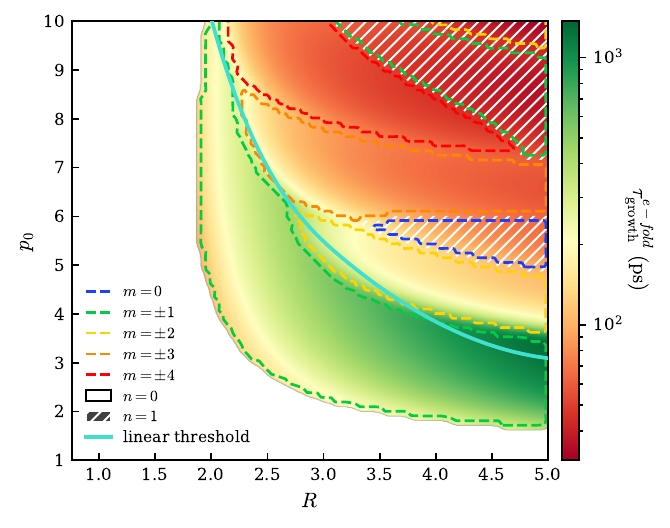}
\caption{\textbf{Growth time map of the compact self-trapped solution.}
$e$-folding growth time $\tau_{\rm growth}^{e\text{-}fold} = \Gamma^{-1}/\mathrm{Im}(\omega)$
of the dominant BdG instability on the $(R,p_0)$ plane, for $\Gamma^{-1}=100\,\mathrm{ps}$
(log color scale). Dashed lines: stability boundaries for $m=0,\pm1,\ldots,\pm4$,
as in Fig.~4a; hatching: region where the dominant mode has $n=1$ rather than
$n=0$. Cyan: linear condensation threshold. $R$ and $p_0$ are dimensionless.
Parameters: $\varepsilon=8$, $\xi=1$, $a=1$.}
\label{fig:S4}
\end{figure}

The weak instability discussed in the Main Text and End Matter was illustrated for a single pump spot radius, $R=4$, where the dominant $m=1$, $n=0$ BdG mode gives $\tau_{\rm growth}^{e\text{-}fold} \approx 1.5\,\mathrm{ns}$
for $\Gamma^{-1}=100\,\mathrm{ps}$. Figure~S3 extends this calculation over the full $(R,p_0)$ existence domain of the compact branch, confirming that this is a generic feature rather than a property of one point: below and
near the linear threshold, growth times of hundreds of ps to a few ns persist throughout the existence region, one to two orders of magnitude longer than $\Gamma^{-1}$. Above threshold, as $p_0$ increases the dominant unstable mode
shifts to progressively higher $m$ (and eventually to higher $n$), and the growth time drops sharply, down to a few ps at the largest pump powers considered. This confirms that the sub-threshold self-trapped state is not merely a
theoretical curiosity confined to a fine-tuned point, but remains observable on nanosecond timescales across its entire domain of existence, opening a realistic window for its direct experimental observation under steady-state (CW) pumping conditions.

\end{document}